# Discussion of Features for Acoustic Anomaly Detection under Industrial Disturbing Noise in an End-of-Line Test of Geared Motors

Peter Wißbrock
*Innovation*
*Lenze SE*
Aerzen, Germany
peter.wissbrock@lenze.com

David Pelkmann
*Institut für Systemdynamik und Mechatronik*
*Bielefeld University of Applied Sciences*
Bielefeld, Germany
david.pelkmann@fh-bielefeld.de

Yvonne Richter
*Institut für Systemdynamik und Mechatronik*
*Bielefeld University of Applied Sciences*
Bielefeld, Germany
yvonne.richter@fh-bielefeld.de

*Abstract—* In the end-of-line test of geared motors, the evaluation of product quality is important. Due to time constraints and the high diversity of variants, acoustic measurements are more economical than vibration measurements. However, the acoustic data is affected by industrial disturbing noise. Therefore, the aim of this study is to investigate the robustness of features used for anomaly detection in geared motor end-of-line testing. A real-world dataset with typical faults and acoustic disturbances is recorded by an acoustic array. This includes industrial noise from the production and systematically produced disturbances, used to compare the robustness. Overall, it is proposed to apply features extracted from a log-envelope spectrum together with psychoacoustic features. The anomaly detection is done by using the isolation forest or the more universal bagging random miner. Most disturbances can be circumvented, while the use of a hammer or air pressure often causes problems. In general, these results are important for condition monitoring tasks that are based on acoustic or vibration measurements. Furthermore, a real-world problem description is presented to improve common signal processing and machine learning tasks.

## I. Introduction

Industrial geared motors are often complex and freely or modularly configured components. This leads to a high diversity of variants and a decreasing batch size. Therefore, it is hard for the assembly worker to evaluate the product quality. An acoustic-based and automated end-of-line test is used to evaluate, whether a product is of high quality or not. This is difficult in an industrial environment with acoustic disturbances. The measurement system must match several requirements. To achieve cost optimization, the time to prepare a motor for measurements must be as short as possible. At the same time, the test must be reproducible. Additionally, the frequency range of the sensor must match the fault signature. Common vibration measurements have problems with the connection of the sensor head, the frequency range, or exceed cost limits. However, quality issues reported by a customer are often associated with sound quality. In a previous paper [1], a concept is presented on how an end-of-line test for high varied geared motors can be applied. It is pointed out, that acoustic signals should be used instead, which can often be used in the same way as vibration signals. To set up the motor for measurement, it is only necessary to position it under the microphone. High-frequency ranges can be used with nearly each microphone. Therefore, acoustic measurements are an obvious choice. However, the authors found out that the main problem in using acoustic signals is the rough acoustic environment in a typical production hall.

Common end-of-line tests work for minor or major series production with machine learning. Due to the high diversity of variants and small batch sizes, constructing adequately sized datasets is challenging. To the best of the authors' knowledge, no fully labeled real-world datasets are available for the described scenario. It is realistic to initially collect only data that does not have significant errors. However, no label or quantification exists for minor faults or environmental disturbances and in complex systems, not all types of faults are known. Semi-supervised anomaly detection can be used to circumvent the problem of missing labeled datasets. However, the influence of disturbances on common anomaly detection methods is unknown. Many common machine learning methods are feature-based, and it is often discussed, which are the best features to be used. But it is rarely shown, how to handle changing environmental conditions like noise or vibration, which does not come from the device under test. Additionally, authors often discuss features of only one of the domains signal processing, artificial intelligence, or psychoacoustics. These research gaps make it difficult to decide, which features or if any should be used.

To address the above challenges, a measuring system is described, which is used to record an acoustic real-world dataset. One type of geared motor with different fault characteristics and disturbing noises are included. This paper compares the robustness of commonly used types of features and anomaly detection approaches. Data from healthy geared motors are used to train a system, which identifies whether a geared motor is in fault condition. Faulty examples and those including acoustic disturbances are used for evaluation.

## II. Related Work

A lot of research is done in the field of motor inspection. In principle, condition monitoring tasks consider similar types of faults, and thus the same approaches could be useful. However, faults that occur in quality inspection tasks have different intensities and severities. Therefore, literature is presented focusing either on end-of-line quality inspection or on disturbances in measured data.

### A. End-of-Line Quality Inspection of Gears and Motors

Related work in quality inspection discusses various combinations of features and machine learning methods.

Parts of this article were supported by the Ministry of Economic Affairs, Innovation, Digitization, and Energy of North Rhine Westphalia through the excellence cluster itsOWL in the projects "PsyMe" and "ML4Pro²".

Logarithm-scaled acoustical signals are used for statistical evaluation and regression of vehicle-gear quality [2]. Acoustical center frequencies of DC-motors are taken as input for a multi-layer perceptron (MLP) classification [3]. A vibration-cepstrogram of motors is taken as input for an autoencoder (AE) based on deep-belief networks as a semi-supervised task [4]. Then, a validation set is used to choose a feasible fault-threshold. However, psychoacoustic features outperform statistical features as input to a simple MLP classification [5]. Statistical features of time batches of a positioning signal are taken as input for an ensemble of one-class-classifiers (OCC) [6]. One classifier is used for each type of fault and the good condition in the quality inspection of steering systems. Log-mel-coefficients of vibration signals serve as input for different architectures of AE in the quality inspection of motors [7]. MLP-AE outperforms other architectures of the AE. The fault-threshold is considered separately for each time slot. To summarize, various statistical and other low-level features dominate in the field of quality inspection, but comparisons to further approaches used in the field of condition monitoring are missing.

### B. Disturbances in Acoustic or Vibration Data

In the field of condition monitoring, related papers mainly discuss three different types of disturbances in acoustic or vibration data leading to different solution approaches. These are impulsive noise, Gaussian noise, and industrial noise. Other parts of a machine or production line may transmit disturbing impulses (impulsive noise), which can be equidistant or randomly distributed over time. These disturbances are often treated with features in the signal processing domain via envelope-demodulation. The log-envelope spectrum is more robust in the presence of impulsive disturbances than the squared envelope [8]. In [9] the data is synchronized to the exactly measured position of the motor, which requires a feedback system. The envelope transformation has the bandpass range as one of its main parameters. A broad comparison for the selection of the informative-frequency band is presented in [10] to find a bandpass range that serves a specific type of fault best. This shows that frequency band selectors generally react to disturbances. Approaches of vibration analysis work for bearing and gear faults [11], [12]. A log-envelope spectrogram is taken in [13], for showing up with features for small frequency ranges, not using a bandpass filter. Often the robustness of neural-network-based approaches for fault classification is tested by adding Gaussian noise to the existing dataset. The anti-noise ability of a convolutional neural network (CNN) is used in [14], [15] for the extraction of robust features. Other approaches apply denoising autoencoders trained directly on corrupted data. An existing model with different noise levels for transfer learning is adapted in [16]. In [17], first the noise is reduced with a one-dimensional denoising CNN-AE and then a CNN is used for classification. The popular MIMII dataset [18], [19] for anomaly detection on acoustic data includes disturbances observed in the industry (industrial noise). A review work [20] shows that papers working on this dataset mostly consider mel-filter cepstral coefficients, known as MFCC or log-mel energies as features for AE, as well as CNN, CNN-AE, or transfer learning methods of pre-trained deep networks.

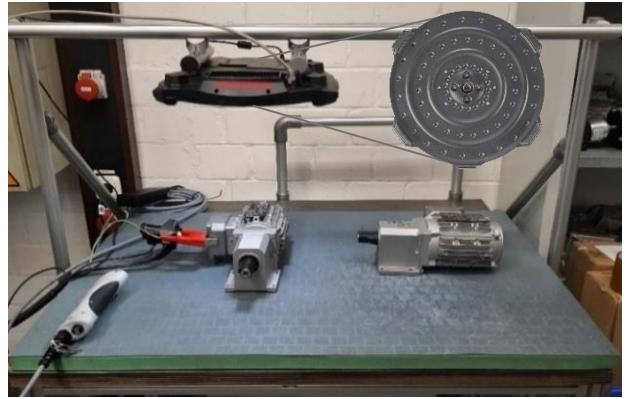

Fig. 1. Acoustic end-of-line test system for geared motors including the microphone array (on top), the device under test (left motor), and vibration decoupling (green base).

## III. EXPERIMENTAL SETUP

A test system is required, which provides a reproducible test run. The test assumes steady operation, driving the motor at nominal speed for several seconds in both clockwise (cw) and counterclockwise (ccw) directions. During the inspection of geared motors with a high gear ratio, the test should include a few full cycles of the slowest shaft. The sensor measurement starts after the motor has reached nominal speed.

### A. Measuring Equipment

Microphones with directive efficiency can be used to reduce the problem of acoustic disturbances. Array microphones with the calculation of the local sound are of rising interest in production environments [21]. The amount of the directive efficiency is associated with the width of the array and works better within high-frequency ranges. Here, an array microphone with micro-electro-mechanical sensors (MEMS) of type SoundCam 2.0 is used. The mounting and the positioning of the motor can be seen in Fig. 1. 64 MEMS are included in the SoundCam, which calculates a local sound with a sampling frequency of 50 kHz. It is placed above an assembly table that is separated from the other parts of the production line. A decoupling base is used to avoid super-positioning of motor vibrations and others coming from the environment. The distance between the microphone array and the base must be chosen. With 400 mm distance, the longest geared motor produced on the specific production line, matches the observed area of the array in a 90° beamwidth-angle. The number of stored sampling values is fixed to $2^{18}$ each for the directions cw and ccw (~5.24s).

### B. Tested Devices

All tested geared motors are of the same type. It is a 50 Hz three-phase induction motor with a rated velocity of 1375 rpm and 90 W power. There is no ventilation unit or feedback system. The gear is a two-stage helical one with a gear ratio of 10. The dataset includes measurements of three subsets. The first includes tests directly in a production line, while the second includes a set of motors that have been discarded due to quality issues. Those motors have been tested with the same end-of-line test system, but in a quality inspection area right next to the production, where regularly lower environmental noises can be expected. The third systematically describes acoustic disturbances. As the number of motors in this scenario is limited, both directions are handled as separate measurements. This is possible as gears are helical, other types of gears may not have the same acoustic behavior in both directions. In general, different types of faults occur and it is usually not known which one. As experts for quality inspection mentioned, several of the observed faults concern the four toothed wheels but they are not limited to them. In manual evaluation, the acoustic behavior of faults includes different clusters. Some are impulsive, which is typical for pitting at toothed wheels. However, others sound more circumferential.

All samples of the subsets are labeled by evaluating the sound quality. Therefore, all three authors compared the acoustic behavior according to the criteria of general sound quality, comparability to a reference as well as their expert knowledge of faults in geared motors. The rigor of the evaluation corresponds to the objective of motors with special requirements for acoustic quality. As all these evaluations are partly subjective, some fluctuations in labeling are expected. Therefore, the quality is divided into three classes, referred to as 'good', 'warning', and 'error'. The 'good' ones are supposed to pass the test and the 'error' ones do not pass the test. Motors labeled with 'warning' cannot be assigned with certainty to one of the other classes and may include minor quality issues. In practice, these motors would be checked once more by the worker or quality inspection team. In the case of standard acoustic requirements, these would be considered as 'good'. The authors assume that there is overlap in the true classes. Some false classifications will be accepted between the classes of 'good' and 'warning' or 'warning' and 'error'.

### C. Acoustic Disturbances

Acoustic noises, which will always be present in industrial production environments can negatively affect the end-of-line test. These acoustic disturbances may have different characteristics e.g., impulsive or continuous and with specific or broad frequency range. Sources of disturbances include the use of tools like hammering, air pressure, or electric wrenches as well as human speech, music, or ventilation units. Mapping these to the categories observed in literature is not directly possible. Both, impulsive and Gaussian noise occurs in general.

The dataset recorded in production already includes a broad spectrum of acoustic disturbances. However, more specific labeling of different types of disturbances is needed for rating the approached methods. Therefore, an additional subset is created using two of the already seen 'good' and two of the erroneous motors. For each of them, some additional measurements without disturbances and with the following are taken: Using a soft-head hammer on a metal plate, cleaning a mechanical system with short bursts of air pressure, playing pop music, speech in the form of a technical description, activating a 230 Volt-AC ventilation unit, or using an electric wrench with a torque limiter. The strength of all the disturbances is chosen so that they are highly but realistically.

TABLE I. OVERVIEW OF RECORDED DATA

| Subset of Data | No of Occurrences of the Fault Classes | | | Strength of the Noises | | |
|---|---|---|---|---|---|---|
| | good | warning | error | no | low | loud |
| Production Line | 19 | 17 | 4 | 4 | 17 | 19 |
| Quality Inspection | 12 | 8 | 18 | 32 | 3 | 3 |
| Acoustic Disturbances | 30 | 15 | 15 | 11 | 6 | 43 |

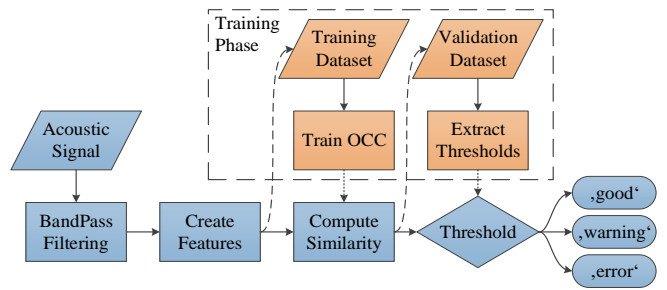

Fig. 2. The general process of end-of-line testing (blue) and systems training (orange)

### IV. APPROACHED METHODS

As shown in Fig. 2, the general approach covers first calculating features on the filtered acoustic data. Then, the anomaly detection method is trained on a training dataset to find a description of samples' similarity using a one-class-classifier (OCC). Using a validation dataset, two feasible thresholds for classifying the data into the three degrees of faults are extracted. The training dataset includes the label 'good' 25 times and 'warning' 17 times. The validation dataset includes all other samples except the disturbance dataset. The robustness of the approached methods is then compared using the disturbance dataset.

### A. Bandpass Filtering

Acoustic arrays provide damping better for higher frequencies e.g., SoundCam half-power-beamwidth of 90° is at 800 Hz. Between 4800 and 5400 Hz, the maximum side-lobe level increases significantly. Bandpass filtering helps to remove disturbances included in the acoustic data or is a fixed part of feature creation. The cutoff frequencies must be chosen, so that acoustic disturbance is avoided, while frequencies stimulated by faults are covered simultaneously. The impulsive frequency range of 'error'-labeled data is approximately between 850 and 5100 Hz. In [11], a similar range of gear fault impulses, 800 to 5000 Hz, is observed so that, higher frequencies than 5100 Hz must not be considered. Considering the subset of the production line, 1050 Hz and less must be excluded regarding disturbances, which can be achieved with a lower cutoff frequency of approximately 1150 Hz. To obtain a filter without phase distortion, a FIR filter with hamming window function is used twice, once forward and backward. The window length is defined, so that it covers 7.5 periods of the lower cutoff frequency. The impact of the bandpass filter is shown in Fig. 3. An alternative approach would be to first reduce disturbances using a denoising CNN. Reproducing the approach of [17] leads to a reduction of the higher frequencies, however, not the lower frequencies, which is mandatory.

### B. Definition OF FEATURES

The evaluated methods include four types of features from different domains. The domain of signal processing covers the log-envelope and the resulting features. Additionally, the domain of time-frequency provides spectral features like log-envelope-spectrum as well as log-mel-coefficients. The mel-scale transforms the frequencies into a scaling as people experience it. The same experience is covered by the domain of psychoacoustics. The fourth domain is representation learning, where features are extracted from a pre-trained CNN.

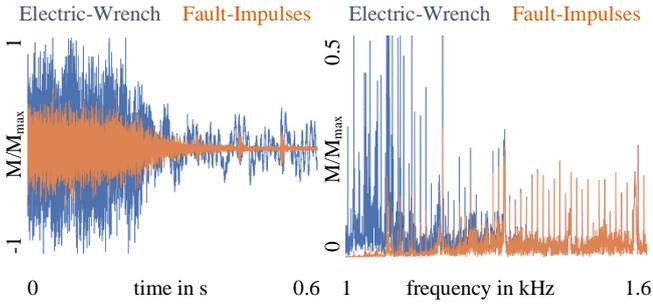

Fig. 3. Normalized magnitudes M of an 'error'-signal before (blue) and after (orange) bandpass filtering. The signal includes the use of an electric wrench for half the time. In the time domain (left), fault impulses are visible after filtering only. Most of the disturbance is suppressed in the frequency domain (right).

*1) Log-Envelope:* Following [8], the log-envelope spectrum (LES) is a variation of the commonly used squared envelope spectrum. Let $\{X[n]\}_{L,H}$ be the analytic signal using the Hilbert-transform of the bandpass filtered acoustic time signal, where L and H are lower and higher cutoff frequencies. The LES for positive frequencies is then defined by applying the following steps (1). Compute the squared absolute, perform log-transform, perform real Fourier-transform, and compute the squared absolute again. These steps lead to a demodulation of the impulses caused by the faults, which stimulate the structural resonances. Considering the disturbances included in the production line dataset, very high contamination of the lowest frequencies of just a few Hz is noticed. In general, higher frequencies provide lower amplitudes, so it is worse to evaluate faults at very high frequencies as well. Taking these observations into account with the LES-features, the used features $LES_{limited}$ are defined for $\mathcal{D}_{LES}$ = [10, 555.5] Hz.

$$LES[f] = \left| \mathcal{F} \left\{ \log \left( \left| \{X[n]\}_{L,H} \right|^2 \right) \right\} \right|^2 \quad (1)$$

*2) LES-Features:* Features extracted from LES at discrete frequency positions are often considered as fault frequencies (FF). The amplitudes at these specific frequencies are then the features $LES_{FF}$. As these frequencies are proportional to the motor velocity, which exact value is unknown, the calculation is provided for rated values. The features $LES_{FF}$ are computed as the maximum amplitudes within a range of FF ± 1%, while 1% is the maximum speed deviation specified for this motor type. If none of the discrete values lies in this range, the nearest one is taken. For faults at toothed wheels, the most important frequencies include the rotational frequencies of the three shafts and their harmonics $FF_s$ (2) as well as the two gear mesh frequencies and their sidebands $FF_g$ (3). Both equations consider three harmonics or sidebands and frequencies in the range of $LES_{limited}$ only. Overall, 32 features are extracted. The highest considered frequency is then the third upper sideband of the first gear mesh and first shaft with 555.5 Hz, which already includes the 1% of tolerance. Other types of faults are not known for the considered dataset but may be included. In general, possible faults include bearing faults, where these FF exist but cannot be calculated for all the bearings included in the gear. These frequencies are lower than the considered upper limit. Other faults like eccentricity shafts can be identified with the already considered $FF_s$.

$$FF_s = i \cdot f_n \ \forall \ i \in \{1,2,3\}; n \in \{1,2\} \quad (2)$$

$$FF_g = f_n \cdot z_{2n-1} \pm i \cdot f_{n+m-1} \forall i \in \{0,1,...\}; n, m \in \{1,2\} \quad (3)$$

*3) Log-Mel Spectrogram:* Log-mel-spectrograms (LMS) are of rising interest to represent the behavior of signals in industrial environments, they represent the overall sound quality. In this approach, data is transformed into the time-frequency domain using short-time-Fourier-transform and applying scaling and reduction techniques to create the features. These are: compute the spectrogram, reduce the number of frequency-bins using filters, and transform in logarithm style. The filters are overlapping triangle windows covering a constant frequency range for normal log-spectrograms. In mel-scale, the frequency range of filters is rising for higher frequencies corresponding to the human hearing. The lowest and highest frequencies for the filters are defined to be 1150 and 5100 Hz respectively, so explicit bandpass filtering is not required. In [22], the performance increases with up to 12 filters, while more filters provide the same performance. As computational cost is not in focus, this boundary is buffered by factor two, so that 24 filters are used, which is a common amount [7]. To ensure that each time slot can contain any fault, the length of the time window is set to four full cycles of the slowest shaft, in this case 1.75s. Time slots are overlapping with 50 %.

*4) Log-Envelope Spectrogram:* An envelope is created from a spectrogram to represent the mechanical faults similar to LES. The basic log-spectrogram is used with a much smaller time window, while other parameters are kept. The length of time windows is chosen so that a local fault of the fastest shaft would be separated, e.g., less than a quarter of a motor's cycle. A second condition is the highest frequency to be observed in the envelope spectrogram, which is given by the time window as period-length. The first 10 Hz are declined as the same phenomenon from disturbances occurs as with the LES, which excludes the harmonics of the first shaft. Therefore, the first gear mesh frequency inclusively 1% of tolerance should be included, which leads to a window length of 0.0088 s for 113 Hz. The envelope-spectrogram is calculated by applying type-two discrete cosine transform to the time domain and taking its amplitude. Max-pooling is applied to the envelope-spectrum domain to reduce the dimensionality and thus compute the features $LES_{spectral}$. A reduction factor of 11 is feasible to keep all possible FF separated.

*5) Psychoacoustics:* To address the sound quality itself, psychoacoustic metrics are discussed, which provide a basis for simulating human hearing. Here the target of psychoacoustics is to generate features that match the worker's "ear"-evaluation as well as possible. The most important psychoacoustic metrics are loudness, sharpness, roughness, and fluctuation strength [24]. After applying the bandpass filter, these four are computed, using the functions which are integrated in MATLAB. In addition, the roughness is computed based on a different definition [25], using

MOSQITO[1]. As a data exploration shows, the stationary loudness and sharpness are more influenced by the disturbances than by the faults. Therefore, only the roughness and fluctuation strength are used as psychoacoustic features *PA*, which are obtained as the root mean square of their time course. These features show better or nearly the same accuracy in classification as statistical features [23]. To obtain a better performance, these features are combined with LES$_{FF}$, referred to as PALFF.

*6) Deep Representation Learning:* The approaches described above require a lot of resources and expertise for feature engineering. This points out the weakness of classical machine learning: the inability to extract and organize different information from raw data. Replacing this manual feature engineering with systems to learn not only prediction or classification, but also data representation is the research area of representation learning. In recent years, many different architectures for deep representation learning have been presented and successfully applied in different domains [26]. In this work, the sample-level CNN with squeeze and excitation block and multi-level feature aggregation architecture (SCSE) is implemented, which has been successfully applied to classify the genre of music [27]. To do so, a larger amount of data is needed than available in this work. Therefore, the *SCSE* is trained with the music dataset Free Music Archive [28]. The input data is resampled to 16 kHz and with a network input size of 59049, the time window is 3.69s. Time slots overlap with 66.67%. Then, the part of the SCSE responsible for learning the representations is extracted and its parameters are fixed. This learned part of the model can then be used to transform raw data into features.

### C. Semi-Supervised Anomaly Detection

Several steps are needed for anomaly detection. First preprocessing, such as scaling or dimensionality reduction is applied to some of the features. Then a machine learning algorithm is trained to provide a score for the similarity between the samples using the training dataset. Therefore, different OCC are compared. At least, the validation dataset is used to extract thresholds for classification as 'good', 'warning', and 'error'.

*1) Preprocessing:* In the usage of features, that grants no information in relation to each other, a scaling transformation is common. Therefore, scaling is performed for the features LES$_{FF}$ and PA. Since there is no guarantee that the training data will be free of contamination, a robust-scaler is supposed to reach better performance compared to a standard-scaler. The scaler removes the median and scales the data of the interquartile range to be in the interval [-1, 1]. For some of the higher-dimensional sets of features, a dimensionality reduction by principal component analysis (PCA) boosts the performance. The training dataset is used to find a set of components covering most of the feature's variance. The number of components is chosen so that 90 % of the variance is kept. This approach benefits the features LMS and SCSE.

*2) One-Class-Classifier:* Two well-known OCC, the one-class support vector machine (OC-SVM) and isolation forest (I-Forest) are compared to the bagging random miner (BRM).

[1] Green Forge Coop. MOSQITO. https://doi.org/10.5281/zenodo.5284054

The OC-SVM learns a decision function on the training data to describe samples' similarity using a rbf-kernel. As input parameter, a fraction of contamination of training data is needed. A comparison of different values is presented within the experimental results, as the true contamination is unknown. The I-Forest creates decision thresholds randomly on randomly chosen features and thus an ensemble of decision trees is created. A samples' similarity is scored as the mean of needed depth to terminate the single trees. A broad comparison of OCC for different data domains [29] found out that the BRM provides the best overall performances and often outperforms OC-SVM and I-Forest. BRM is creating weak classifiers on randomly chosen samples and combines them into an ensemble [30]. The mean similarity is scored based on the nearest neighbor and the distance within a weak classifier. Both I-Forest and BRM have the number of classifiers in the ensemble as a parameter, by default set to 100. Spectrogram representations like LMS are often used in combination with autoencoders (AE), this does not work for the given dataset. Reproducing the architecture in [7], with this dataset, the AE always shows the same output regardless of the input. Using ReLu activation instead of tanh for all except the last layer, this problem is eliminated, however, the AE does not converge to a feasible representation at all.

*3) Thresholding:* Often, evaluation of possible thresholds is shown using receiver operating characteristic (ROC). For different thresholds, the false positive rate shows the proportion of missed faults (MF), and the true positive rate shows the proportion of whether no pseudo-fault (PF) occurs. The faulty data is well separated if the point (0,1) is included in ROC for some threshold. The claimed area under curve (AUC) is used to compare the performance of different approaches, which is one if ROC includes (0,1). Fig. 4 shows the two ROC for the thresholds, which are used to cover all 'good' as ROC$_g$ as well as to cover all 'good' and all 'warning' as ROC$_w$. If the data is not well separated, a feasible method for extracting thresholds is needed. Each of the OCC has a built-in function to calculate a threshold to evaluate, whether a sample belongs to the distribution of training data, using the contamination parameter. However, adding tolerance or an additional threshold for the second degree of fault is not supported.

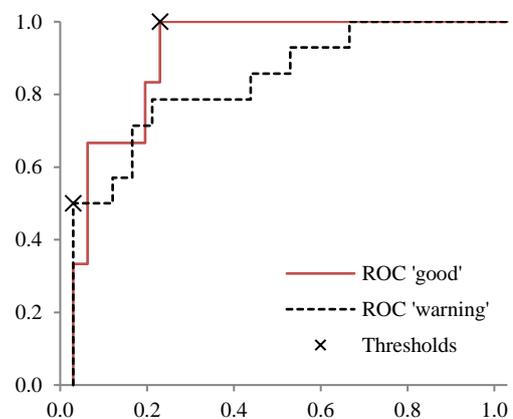

Fig. 4. ROC for one execution of I-Forest with features PALFF for acceptance of 'good' and 'good' joint with 'warning'.

TABLE II. RESULTS FOR VALIDATION DATASET WITH LOWEST AND HIGHEST PERFORMANCE OF FIVE RUNS

| Features | OC-SVM | | | | | | I-Forest | | | | | | BRM | | | | | |
|---|---|---|---|---|---|---|---|---|---|---|---|---|---|---|---|---|---|---|
| | $AUC_g$ | | $AUC_w$ | | $MF_v$ | | $AUC_g$ | | $AUC_w$ | | $MF_v$ | | $AUC_g$ | | $AUC_w$ | | $MF_v$ | |
| | min | max | min | max | max | min | min | max | min | max | max | min | min | max | min | max | max | min |
| $LES_{limited}$ | 74 | 75 | **87** | **88** | **0** | **0** | 54 | 66 | 46 | 57 | 14 | 7 | 70 | 70 | 77 | 77 | 4 | 3 |
| $LES_{FF}$ | 86 | 86 | 80 | 81 | 4 | 2 | **86** | **93** | 78 | 83 | 2 | 0 | 87 | 87 | 82 | 82 | 4 | 3 |
| $LMS_{PCA}$ | 31 | 43 | 45 | 61 | 14 | 6 | 36 | 42 | 48 | 55 | 10 | 7 | 18 | 20 | 41 | 42 | 17 | 15 |
| $LES_{spectral}$ | 85 | 86 | 85 | 86 | 2 | 1 | 55 | 73 | 62 | 85 | 8 | 1 | 83 | 83 | 84 | 85 | 2 | 2 |
| PA | 61 | 66 | 72 | 76 | 6 | 5 | 68 | 72 | 75 | 79 | 4 | 3 | 73 | 75 | 79 | 80 | 3 | 3 |
| PALFF | 86 | 88 | 85 | 86 | 2 | 2 | **89** | **92** | 81 | 84 | **0** | **0** | **89** | **90** | **89** | **89** | **0** | **0** |
| $SCSE_{PCA}$ | 35 | 65 | 47 | 71 | 12 | 5 | 36 | 45 | 53 | 61 | 10 | 7 | 56 | 57 | 71 | 72 | 4 | 3 |

TABLE III. RESULTS FOR DISTURBANCE DATASET FOR THE BEST RUN OF VALIDATION AND FOR MEAN OF FIVE RUNS

| Features | OC-SVM | | | | | | I-Forest | | | | | | BRM | | | | | |
|---|---|---|---|---|---|---|---|---|---|---|---|---|---|---|---|---|---|---|
| | $MF_t$ | | $PF_t$ | | Acc | | $MF_t$ | | $PF_t$ | | Acc | | $MF_t$ | | $PF_t$ | | Acc | |
| | min | mean | min | mean | max | mean | min | mean | min | mean | max | mean | min | mean | min | mean | max | mean |
| $LES_{limited}$ | 1 | 1.0 | **0** | **0.0** | 40 | 42 | 8 | 8.4 | 1 | 4.6 | 33 | 34 | 2 | 2.0 | **0** | **0.0** | 58 | 58 |
| $LES_{FF}$ | 4 | 4.0 | **0** | **0.0** | **70** | **71** | 4 | 3.6 | **0** | **0.0** | 63 | 67 | 4 | 4.0 | **0** | **0.0** | **70** | **70** |
| $LMS_{PCA}$ | 3 | 5.8 | 12 | 12.0 | 28 | 36 | 4 | 4.8 | 12 | 12.0 | 35 | 40 | 7 | 6.4 | 12 | 12.0 | 36 | 36 |
| $LES_{spectral}$ | 1 | 0.8 | **0** | **0.0** | 43 | 43 | 1 | 2.8 | 0 | 1.2 | 23 | 41 | 1 | 1.0 | **0** | **0.0** | 40 | 40 |
| PA | 1 | 1.8 | 5 | 5.4 | 50 | 54 | 1 | 1.0 | 3 | 3.8 | 50 | 48 | 1 | 1.0 | 5 | 5.0 | 55 | 53 |
| PALFF | 4 | 3.2 | 5 | 5.4 | 60 | 61 | 3 | 1.8 | **0** | **0.0** | 66 | 61 | 3 | 3.0 | 2 | 2.0 | 65 | 66 |
| $SCSE_{PCA}$ | 1 | 1.0 | 12 | 12.0 | 46 | 47 | **0** | 0.2 | 12 | 13.0 | 50 | 48 | **0** | **0.0** | 12 | 12.0 | 36 | 36 |

Thresholds are defined using the following algorithm. The threshold considering whether 'error' is assigned or not is $t_e$ and the threshold for classification as 'good' or 'warning' is $t_w$. The presence of PF is worse, considering the requirements of production planning. Therefore, $t_e$ is chosen to describe the border whether exactly one or no PF is included in the validation dataset. Then, $t_w$ is chosen, so that the probability of belonging to class 'good' is constant for prediction and given labels, ignoring 'error' labeled data. The algorithm to extract the thresholds is shown below.

**Algorithm Threshold Detection**

**Input:**
  r: $ROC_g$
  s: scores computed by OCC
  c: true classes of samples in s
**Output:**
  $t_e$: threshold to be in 'error'
  $t_w$: threshold to be in 'warning'
**Steps:**
  1: $t_e$ is the first element of s for that the true positive rate of r is one
  2: eliminate elements in s for that c is 'error'
  3: sort s with increasing similarity
  4: $t_w$ is the n's element of s, where n is the occurrence of 'good' in c

## V. EXPERIMENTAL RESULTS

For each combination of features and OCC, the algorithms run five times. For the OC-SVM the contamination parameter is varied in range of {0.1, 0.2, 0.3, 0.4, 0.5}, while this parameter is not used by the other OCCs if the built-in threshold detection is not used. The runs of I-Forest and BRM are used to evaluate whether these ensembles are dependent on their random state. First, the performance according to the detectability of faults and the separability of datasets is evaluated using the validation dataset, shown in TABLE II. The $AUC_g$ in % of $ROC_g$ describes the performance for only 'good' is accepted and $AUC_w$ in % of $ROC_w$ for 'good' and 'warning' is accepted, respectively. Higher AUC promises better separability of the classes. However, a higher AUC does not guarantee a higher number of detected faults. Therefore, the number of missed faults (MF) is also considered. $MF_v$ shows the number of anomalies in the validation data, that cannot be detected. The influence of contamination or randomness is evaluated showing the range of performance for best and worst model, respectively. For an active system, a model must be selected, and one would choose the model for that $MF_v$ is lowest and if of the same value, AUC is highest. PALFF perform without $MF_v$ as well as $LES_{limited}$, but with significantly better $AUC_g$. With a maximum of two $MF_v$, $LES_{spectral}$ and $LES_{FF}$ are also good choices. Considering the OCCs, it can be observed that OC-SVM perform better for high-dimensional data like $LES_{lim}$ and $LES_{spectral}$ as I-Forest does. While the I-Forest is better on lower dimensional data, it is highly depending on its random state. The BRM is a fair compromise over all features and outperforms others for PA and PALFF. The best combinations of features and OCC are:

1. PALFF and BRM (no $MF_v$ and best $AUC_w$)
2. PALFF and I-Forest (no $MF_v$ and best $AUC_g$)
3. $LES_{limited}$ and OC-SVM (no $MF_v$)
4. $LES_{spectral}$ and OC-SVM
5. $LES_{FF}$ and I-Forest

In the next step, the robustness against acoustic disturbances is evaluated presenting the disturbance dataset to the best of the five models. The performance is shown in TABLE III. along with the mean of all five runs. As the main target of a quality inspection system is to detect faults, the number of MF is also considered on the disturbance set. Thereby, no 'good' samples should be predicted as faulty, referred to as pseudo-faults (PF). $MF_t$ shows the number of anomalies, that cannot be detected, while $PF_t$ is the number of samples predicted as 'error', that are 'good' (lower is better). To guarantee that the values of $MF_t$ and $PF_t$ are trustable, the accuracy (Acc) is considered. As three classes are defined, a random acting approach would achieve an Acc of one-third. An Acc of one is not expected considering the subjectiveness of the labels and the overlapping with class 'warning'. A low Acc in combination with good values for $MF_t$ and $PF_t$ may follow from a system considering most samples as 'warning', which would be as useless as a randomly acting system. Only two types of features reach a feasible Acc, $LES_{FF}$ and PALFF, which includes $LES_{FF}$. While PALFF show a better $MF_t$, $LES_{FF}$ outperforms it with the Acc. For $LES_{FF}$, the OC-SVM or the BRM and for PALFF the I-Forest perform best. Choosing the best model for validation data does not guarantee the best robustness as this list of best choices of the disturbance-dataset demonstrates:

1. $LES_{FF}$ and OC-SVM (best $PF_t$ and Acc)
2. $LES_{FF}$ and BRM (best $PF_t$ and Acc)
3. PALFF and I-Forest (best $MF_t$)
4. PALFF and BRM

Different types of disturbances may lead to PF with different probability. A representative overview is shown in TABLE IV. The classification of 'good' samples is evaluated for the features PA and SCSE, with PF for both cases, 'warning' or 'error' instead of 'good' prediction shown as a sum over all OCCs. For PA most PF occurs with air pressure and the hammer, while speech and the electric wrench cause less trouble. Less PF occurs with music and ventilation. For SCSE air pressure, hammer, and additionally electric wrench lead always into PF. The ventilation unit and sometimes music leads to lower PF. However, for speech SCSE is more robust compared to PA. Hammer and air pressure, followed by electric wrench are most disturbing and with the highest amplitudes at the same time.

TABLE IV.     OVERVIEW OF FALSE ALARMS DUE TO DISTURBANCES

| Disturbance | Occurrences of False Alarm as Sum of All OCCs | | | | | |
|---|---|---|---|---|---|---|
| | PA | | | SCSE | | |
| | good | warning | error | good | warning | error |
| No Noise | 12 | 0 | 0 | 9 | 3 | 0 |
| Speech | 0 | 9 | 0 | 5 | 4 | 0 |
| Music | 10 | 2 | 0 | 7 | 5 | 0 |
| Electric Wrench | 7 | 5 | 0 | 0 | 0 | 12 |
| Ventilation Unit | 12 | 0 | 0 | 2 | 10 | 0 |
| Air Pressure | 0 | 4 | 8 | 0 | 0 | 12 |
| Hammer | 0 | 6 | 6 | 0 | 0 | 12 |

## VI. CONCLUSION

This paper covers the problem of end-of-line testing of geared motors, where only a small dataset can be used. First, a measurement setup is described using an acoustic array. The real-world dataset is labeled by three classes describing the degree of fault. Then different features and one-class-classifier are selected for comparison according to their detectability of faults and robustness against acoustic disturbances. Thereby, the combination of fault frequencies extracted from a log-envelope spectrum and psychoacoustic features (PALFF) shows the best detectability using the bagging random miner (BRM). However, the best robustness is shown when only these fault frequencies ($LES_{FF}$) are used with the one-class support vector machine (OC-SVM). Further features like the log-envelope spectrum, its spectral representation, or the psychoacoustic features also achieve good results in detectability. However, they show some weaknesses in robustness. For anomaly detection, a method should be selected, which fits best the features and requirements. The bagging random miner is the most universal method, while isolation forest and one-class support vector machine outperform it for some features. In the proposed dataset, the use of a hammer or air pressure is the most disturbing noise, while an electric wrench or speech sometimes cause problems. The commonly used approaches such as log-mel-energies and neural networks need particular attention regarding the discussed disturbances.

The next steps of research will include techniques to detect, and if possible, remove disturbing noises on acoustic datasets. This is identified as a mandatory step to use representation learning, which will also be improved in the future, e.g., using a dataset including different variations of geared motors for training the network. Furthermore, it will be discussed how the proposed methods can be transformed into datasets of other variants of geared motors. This is also needed for datasets, that include only good samples. Additionally, the robustness of the one-class-classifier against longer-term changes at the end-of-line test will be considered, using stream-learning approaches. Other tasks will address the further development of the usage of psychoacoustic features for quality inspection of geared motors.